\documentclass[%
 reprint,
superscriptaddress,
 amsmath,amssymb,
 aps,
]{revtex4-2}

\usepackage{dcolumn}
\usepackage{bm}

\usepackage{physics}
\RequirePackage{url}
\RequirePackage[colorlinks=true, allcolors=blue]{hyperref}
\usepackage{graphicx, float}
\usepackage{caption,lipsum}
\usepackage{subcaption}
\usepackage{ragged2e}
\DeclareCaptionJustification{justified}{\justifying}
\begin{document}

\preprint{APS/123-QED}

\title{Quantum optical
coherence microscopy for bioimaging applications}

\author{Pablo Yepiz-Graciano}
\email{pablodgy@gmail.com}
\affiliation{Instituto de Ciencias Nucleares, Universidad Nacional Aut\'{o}noma de M\'{e}xico, CDMX 04510, Mexico}%
\author{Zeferino Ibarra-Borja}
\affiliation{Instituto de Ciencias Nucleares, Universidad Nacional Aut\'{o}noma de M\'{e}xico, CDMX 04510, Mexico}%
\affiliation{Centro de Investigaciones en \'Optica A. C., Le\'{o}n, Guanajuato 37150, Mexico}%

\author{Roberto Ram\'irez Alarc\'on}
\affiliation{Centro de Investigaciones en \'Optica A. C., Le\'{o}n, Guanajuato 37150, Mexico}%

\author{Gerardo Guti\'errez-Torres}
\affiliation{Centro de Investigaciones en \'Optica A. C., Le\'{o}n, Guanajuato 37150, Mexico}%

\author{H\'ector Cruz-Ram\'irez}
\affiliation{Instituto de Ciencias Nucleares, Universidad Nacional Aut\'{o}noma de M\'{e}xico, CDMX 04510, Mexico}%

\author{Dorilian Lopez-Mago}
\affiliation{Tecnologico de Monterrey, Escuela de Ingenier\'{i}a y Ciencias, Monterrey, N.L. 64849, Mexico}

\author{Alfred B. U'Ren}
\email{alfred.uren@correo.nucleares.unam.mx}
\affiliation{Instituto de Ciencias Nucleares, Universidad Nacional Aut\'{o}noma de M\'{e}xico, CDMX 04510, Mexico}%





\begin{abstract}
Quantum-optical coherence tomography (QOCT) is an optical sectioning modality based on the quantum interference of entangled photon pairs in the well-known Hong-Ou-Mandel interferometer.   
Despite its promise, the technique is far from being competitive with current classical OCT devices due to the long required acquisition times, derived from the low photon-pair emission rates. 
In this work, we on the one hand demonstrate a quantum optical coherence \emph{microscopy} (QOCM) technique, based on full-field QOCT, which employs entangled collinear photon pairs in a Linnik interferometer  designed to overcome some of the limitations of previous QOCT implementations, and on the other hand test it on representative samples, including glass layers with manufactured transverse patterns and metal-coated biological specimens.
In our setup, while the idler photon is collected with a multi-mode fiber, the signal photon is detected by an ICCD camera, leading to full-field transverse reconstruction through a single axial acquisition sequence.  
Interestingly, our setup permits concurrent OCT and QOCT trace acquisition, the former with greater counts and the latter with the benefit of quantum-conferred advantages.
We hope that our current results will represent a significant step forward towards the actual applicability of QOCT, e.g. in clinical settings.
\end{abstract}

\maketitle


\section{\label{sec:intro}Introduction}

Optical Coherence Tomography (OCT), reported in 1991 by Huang \textit{et al.} \cite{Huang1991}, permits high resolution, cross-sectional tomographic images of materials and biological samples by measuring back-reflected light.   While it has become an effective technique used in several  branches of medicine (including ophthalmology, dermatology, dentistry, gynecology, cardiology, and laryngology), its use has also been proposed for certain non-biological applications such as the study of polymer matrix composites, as well as for the  retrieval of optical data from multilayer optical disks (see \cite{OCThandbook}).   Although in the span of three decades OCT has to a degree become an established technology, areas of opportunity remain; for example, improving the resolution could permit imaging at an intra-retinal level \cite{Drexler2001}.

The quantum optics community has shown that nonclassical interferometers can be similarly used for optical sectioning, providing some advantages over OCT \cite{Abouraddy2002,Valles2018,MacHado2020}. One such technique is quantum OCT (QOCT), which is the topic of this work \cite{Teich2011}.  QOCT offers on the one hand an enhancement in the axial resolution by a factor of $2$, for a given spectral bandwidth of the light source employed, and on the other hand immunity to dispersion \cite{Abouraddy2002}.   Additionally, it has been noted that the QOCT interferogram visibility is independent of optical losses\cite{Reisner2022}; the  combined resulting immunity to dispersion and losses makes such an approach well suited for unknown biological samples.

These advantages stem from the frequency correlations present in the photon pairs used as the basis for QOCT, which are typically produced by the process of spontaneous parametric downconversion (SPDC). In QOCT, one of the downconverted photons travels a known path length, while its sibling is reflected from the sample, prior to being brought together to interfere at a beam splitter. If they are indistinguishable, they will bunch and exit together through one of the beamsplitter output ports, in an effect known as Hong Ou Mandel (HOM) interference \cite{Hong1987}. A drop in the coincidence detection rate as a function of the signal-idler temporal delay $\tau$ will then appear, centered at $\tau=0$, which is referred to as the HOM dip.  If the sample contains multiple interfaces, a sequence of resulting HOM dips can yield axial morphological information.

The feasibility of QOCT was demonstrated in a proof-of-concept experiment by Nasr et al. \cite{Nasr2003}. It was followed by a series of experiments demonstrating the technique on different samples, such as dispersive media \cite{Nasr2004}, biological specimens \cite{Nasr2009}, and polarization-sensitive materials \cite{Booth2011}; for more information about the initial development of QOCT, see Ref. \cite{Teich2011}. Nevertheless, QOCT has not  become a practical technology, e.g. applicable in clinical settings, due to the need for long acquisition times derived from the low generation rates.

In order to overcome the drawbacks mentioned above, we have in the past demonstrated a number of strategies. An alternative QOCT configuration, relying on a Michelson two-photon interferometer (instead of a HOM interferogram)  was reported by one of us in  \cite{Lopez-Mago2012b,Lopez-Mago2012e}. Notably, this configuration uses a collinear photon-pair source which facilitates optical alignment and leads to a much greater usable flux since in principle all photon pairs produced can  be used, in a comparatively compact and robust system.    We have additionally demonstrated a quantum engineering recipe based on broadband light for the removal of unwanted signal features (artifacts) resulting from quantum interference effects \cite{Graciano2019}, as well as Spectral-Domain QOCT (SD-QOCT) which exploits spectrally resolved measurements to suppress the need for axial scanning \cite{Yepiz-Graciano2020} (see also \cite{Kolenderska2020}).  Finally, while most realistic applications require 3D information about the sample (transverse and axial), a typical QOCT apparatus requires an independent axial scan for each transverse position leading to prohibitively long acquisition times.  With this in mind,  we recently developed Full-Field QOCT (FF-QOCT) \cite{Ibarra-Borja2020}, which permits full 3D sample reconstruction through a single axial acquisition sequence.

It is our aim in this contribution to develop an optimized QOCT apparatus in terms of: i) detection rate, ii) axial resolution, iii) extent of transverse detection window in the sample, and iv) transverse resolution.  We hereby propose and experimentally demonstrate an improved QOCT platform, which we henceforth refer to as Quantum-Optical Coherence Microscopy (QOCM). This platform incorporates the use of: i) a collinear SPDC source, ii) the use of a two-photon \emph{Linnik interferometer} (instead of a HOM interferometer), iii) multi-mode detection for the idler photon / full field detection for the signal photon, and iv) an SPDC pump linewidth  which is sufficiently large for the nearly full suppression of artifacts due to cross-interference \cite{Graciano2019}.   The use of a collinear source implies that in principle the entire photon-pair stream produced at the source may be used in the experiment, without the need for spatial filtering to ensure high-visibility interference (as would indeed be the case for a non-collinear version of the experiment --- see for example Ref. \cite{Ibarra-Borja2020}). 

The photon-pair stream from such a source constitutes the input to a Linnik interferometer,  a variant of a Michelson interferometer in which a microscope objective is included in the sample arm --- resulting in a microscopy device which can probe the sample with a $\mu$m transverse resolution, while retaining all the quantum-conferred  advantages of QOCT.   Note that the use of multi-mode idler detection (i.e. coupling into multi-mode fiber prior to detection) has two advantages: i) it boosts the collected flux, and ii) the resulting weak idler-photon spatial filtering enlarges the extent of the signal-photon transverse detection window. Additionally, it has been reported that such multi-mode detection can permit a higher spatial resolution \cite{Devaux2020}.  Interestingly, because the core of our QOCM setup is identical to that of a classical OCT, by ignoring the photon-pair nature of our light source, i.e. by focusing on the single-channel counts,  we are able to obtain a full OCT measurement concurrently with each QOCT measurement.

Thus, based on our previous experience,  in this work we elevate the capabilities of QOCT one step closer to practical bio-imaging applicability (e.g. in clinical settings). Our QOCM apparatus  combines the 3D imaging capabilities of FF-QOCT based on a time-gated ICCD camera \cite{Ibarra-Borja2020} with the robustness, simplicity, compactness, and increased photon flux offered by a two-photon Linnik interferometer, in a design which nearly fully suppresses cross interference effects.    Here we report a proof-of-concept experiment based on both synthetic samples (copper-coated glass cover slip with engraved designs on both surfaces), as well as biological samples (metal-coated onion tissue and dragonfly wing). We additionally describe an extended theory of collinear-source QOCT with arbitrary signal-idler spectral correlations which permits us to identify experimental parameters leading to artifact suppression.

The content of this article is organized as follows. In section 2 we establish the general theory of QOCT with collinear entangled photons assuming arbitrary signal-idler correlations. Section 3 describes the experiment and methods. Section 4 shows our experimental results. We conclude in section 5.


\section{Theory of QOCT with collinear entangled photons}  \label{sec:theo}

In a previous work, we discussed the theory of QOCT based on a two-photon Michelson interferometer \cite{Lopez-Mago2012b}.  In this paper we present experimental results obtained with a QOCT apparatus based on a Linnik interferometer, which differs from a Michelson in the presence of a microscope objective in the sample arm. The theory in this work, to be shown in this section, pertains to QOCT based on either a Michelson or Linnik interferometer, in which we allow for arbitrary spectral correlations in the photon-pair state.

\begin{figure*}
	 \centering
	  \includegraphics[width=13 cm]{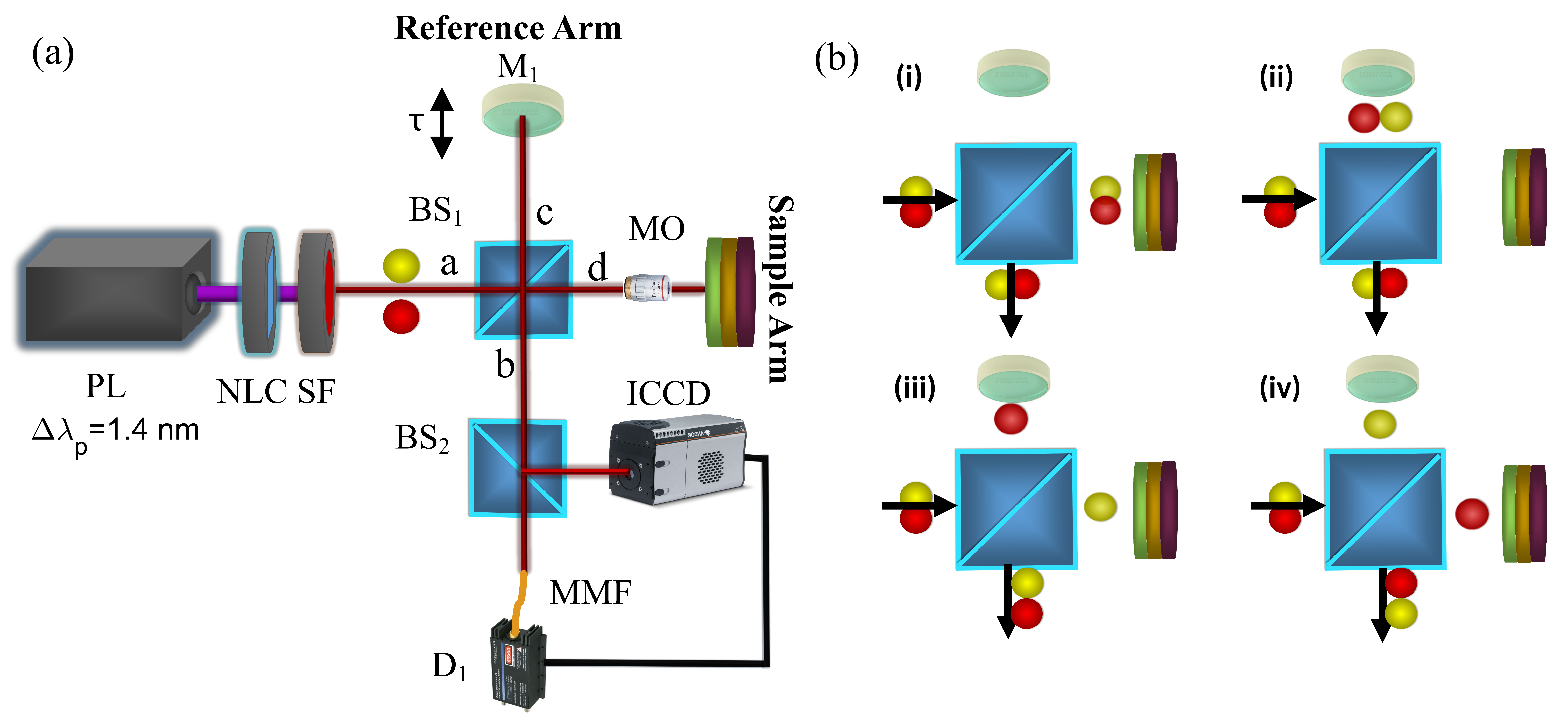}
	   \caption{(a) Schematic of QOCT relying on collinear photon pairs in a Michelson interferometer. (b) Illustration of the interfering pathways at the beamsplitter, post-selected so that both photons exit in the $\hat{b}$ optical mode. PL: pump laser, NLC: nonlinear crystal, SF: spectral filter, BS: beam splitter, M: mirror, D: detector, CC: coincidences counter.}
	   \label{fig:schematic}
\end{figure*}

Figure \ref{fig:schematic}(a) shows a simplified schematic of the two-photon Michelson interferometer. The first component is the  photon-pair source which has as important attributes collinearity and frequency-degeneracy; our experiments are based on a type-I SPDC source.    The quantum state which results from this process may be written as

\begin{equation}\label{Eq:Quantum_State}
    \ket{\Psi} = \ket{\mathrm{vac}} + \eta \iint\mathrm{d}\omega_s \mathrm{d}\omega_i f( \omega_s , \omega_i )\hat{a}^{\dagger}_s(\omega_s) \hat{a}^{\dagger}_i(\omega_i)\ket{\mathrm{vac}},
\end{equation} 
where $\eta$ is a constant related to the conversion efficiency, $f( \omega_s , \omega_i )$ is the joint spectral amplitude (JSA) given in terms of the signal (s) and idler (i) fields with angular frequencies $\omega_{s,i}$, and $\hat{a}^{\dagger}_{s,i}$ are the corresponding creation operators. Energy conservation implies that $\omega_s + \omega_i = \omega_p$, where $\omega_p$ is the pump photon angular frequency. We assume a low parametric gain so that multiple-pair generation may be disregarded.

The joint spectral intensity (JSI) given by $S(\omega_s,\omega_i)=|f( \omega_s , \omega_i )|^2$ is normalized so that the integral of $S(\omega_s,\omega_i)$ over all $\omega_s, \omega_i$ yields unity. It can be written in terms of the pump spectral density $P(\omega_p)$, of the phase-matching function $h(\omega_s,\omega_i,\omega_p)$, and of a spectral filter function $F(\omega)$ applied to each of the SPDC photons as follows \cite{OuBook}:
\begin{align}
    S(\omega_s,\omega_i) &=S_0\, P(\omega_s+\omega_i)|h(\omega_s,\omega_i,\omega_s + \omega_i)|^2 \\ \nonumber  
    &\times F(\omega_s)F(\omega_i),
\end{align}
where $S_0$ is a normalization constant, $h(\omega_s,\omega_i,\omega_p)=\mathrm{sinc}(\Delta K L/2)$, with $\Delta K$  the phase mismatch given by
\begin{equation}
    \Delta K=[n_e(\omega_p)\omega_p - n_o(\omega_s)\omega_s - n_o(\omega_i)\omega_i]/c,
\end{equation}
in terms of the ordinary (extraordinary) index of refraction $n_o$ ($n_e$),  and $L$ is the length of the nonlinear crystal (NLC). We consider a broadband Gaussian beam pump with spectral density centered at $\omega_{p0}$ and with width $\sigma$, i.e.,
\begin{equation}
    P(\omega_p) \propto \exp\left[ - \frac{(\omega_p - \omega_{p0})^2}{\sigma^2} \right].
\end{equation}

The second main component is the interferometer. 
The photons from each collinear pair (both described by mode $\hat{a}$) impinge on one of the beamsplitter input ports.   There are four possibilities for the photon pairs to propagate upon exiting the BS (in modes $\hat{c}$ and $\hat{d}$) as illustrated in Fig. \ref{fig:schematic}(b): either both photons travel together [cases (i) and (ii)], or travel apart [cases (iii) and (iv)].    

The derivation of these four possibilities is given by the following input-output relations of the creation operators for the Michelson interferometer:
\begin{equation}
    \begin{pmatrix}
    \hat{a}^{\dagger}_{\mathrm{in}}\\ \hat{b}^{\dagger}_{\mathrm{in}}
    \end{pmatrix}
    = \begin{pmatrix}
    r^2 e^{i\omega \tau} + t^2 &
    rt e^{i\omega \tau} + tr \\ tr e^{i\omega \tau} + rt & t^2 e^{i\omega \tau} + r^2
    \end{pmatrix}
    \begin{pmatrix}
    \hat{a}^{\dagger}_{\mathrm{out}}\\ \hat{b}^{\dagger}_{\mathrm{out}}
    \end{pmatrix},
\end{equation}
where $r,t$ are the reflection and transmission coefficients of the beam splitter (BS), respectively, and $\omega \tau$ is the phase difference between the optical paths. In our simulations we assume a 50:50 symmetric BS, i.e., with reflection coefficient $r=i/\sqrt{2}$ and transmission coefficient $t=1/\sqrt{2}$.

Another important component is the sample placed as end-reflector in one of the interferometer arms. We characterize the sample by  a transfer function $H(\omega)$, expressed to first order in the sample dispersion as
\begin{equation}
    H(\omega)=\sum_{j=0}^{N-1} r^{(j)} e^{i\omega T^{(j)}}=r^{(0)}+r^{(1)}e^{i\omega T^{(1)}}+...,
\end{equation}
where $N$ is the number of interfaces,  $r^{(j)}$ is the reflectivity  of the \textit{j}th interface and $T^{(j)}$ is the time traveled from the zeroth layer to the \textit{j}th interface and back to the first.

We now proceed to derive the output state using the above transformations.  Note that among all possible ways in which the photon pairs may emerge from the second pass of the photon pairs through the BS, we post-select events in which both photons emerge through mode $\hat{b}$.  With this in mind, the post-selected two-photon state may be expressed as
\begin{equation}
\begin{split}
    \ket{\Psi}_{\mathrm{out}} =& \ket{\mathrm{vac}} + \eta \iint\mathrm{d}\omega_s \mathrm{d}\omega_i f( \omega_s , \omega_i ) r^2 t^2 \left( e^{i\omega_s \tau}+ H(\omega_s) \right) \\
    \times&\left( e^{i\omega_i \tau}+H(\omega_i) \right)  \hat{b}^{\dagger}_{s,\mathrm{out}}(\omega_s) \hat{b}^{\dagger}_{i,\mathrm{out}}(\omega_i)\ket{\mathrm{vac}}.
    \end{split}
\end{equation}

The final step is to detect the signal and idler photons emerging from BS$_1$ in coincidence. For this purpose, the interferometer output is split with the help of a second beamsplitter (BS$_2$), with single-photon detectors placed at both output ports. The resulting interferogram is given in terms of the frequency detuning variables $\nu_{s,i}\equiv\omega_{s,i}-\omega_0$ (where $\omega_0\equiv\omega_{p0}/2$ is the degenerate SPDC frequency) by

\begin{equation}\label{eq:interferogram_sample}
\begin{split}
M(\tau)=&\frac{1}{4}\iint \mathrm{d}\nu_s \mathrm{d}\nu_i \, S_{sym}(\nu_s,\nu_i)\times \\
&\abs{e^{i(\omega_0+\nu_s)\tau}+H(\omega_0+\nu_s)}^2 \abs{e^{i(\omega_0+\nu_i) \tau}+H(\omega_0+\nu_i)}^2,
\end{split}
\end{equation}
where the $1/4$ factor comes from considering a 50:50 beamsplitter, and where we note that BS$_2$ results in the symmetrization of the joint spectrum so that

\begin{equation}
S_{sym}(\nu_s,\nu_i)=S(\omega_0+\nu_s,\omega_0+\nu_i)+S(\omega_0+\nu_i,\omega_0+\nu_s)
\end{equation}
appears in Eq. (\ref{eq:interferogram_sample}), instead of $S(\omega_s,\omega_i)$.

We can gain meaningful information about the coincidence interferogram by expanding $M(\tau)$ in Eq. (\ref{eq:interferogram_sample}). We can show that
\begin{eqnarray} \label{Eq:Mtau}
    M(\tau)&=&M_0+2\mathrm{Re}\{M_1(\tau)\}+4\mathrm{Re}\{M_2(\tau)e^{-i\omega_0 \tau}\} \nonumber  \\ 
    & & +2\mathrm{Re}\{M_3 (\tau) e^{-i2\omega_0 \tau} \},
     \label{eq:expansion}
\end{eqnarray}
where $\omega_0=\omega_{p0}/2$. $M_0$ corresponds to the average number of coincidences, i.e.,
\begin{align} 
    M_0 &= \frac{1}{4} \iint \mathrm{d}\nu_s \mathrm{d}\nu_i\, S_{sym}(\nu_s,\nu_i) \nonumber \\
    &\times \left(1+\abs{H(\omega_0+\nu_s)}^2\right)\left(1+\abs{H(\omega_0+\nu_i)}^2\right),
\end{align}

The envelope functions $M_{1}(\tau)$, $M_{2}(\tau)$ and $M_{3}(\tau)$ appearing in Eq. (\ref{eq:expansion}) are given by
\begin{align}\label{eq:qoct_interferogram}
    M_1(\tau)&= \frac{1}{4} \iint \mathrm{d}\nu_s \mathrm{d}\nu_i\, S_{sym}(\nu_s,\nu_i)  \nonumber \\ 
    &\times H(\omega_0+\nu_s)H^{\ast}(\omega_0+\nu_i) e^{-i(\nu_s-\nu_i) \tau},\\
    M_2(\tau)&= \frac{1}{4} \iint \mathrm{d}\nu_s \mathrm{d}\nu_i\, S_{sym}(\nu_s,\nu_i) \nonumber \\
    &\times  \left(1+\abs{H(\omega_0+\nu_s)}^2\right) H(\omega_0+\nu_i)e^{i\nu_s \tau},\\
    M_3(\tau)&= \frac{1}{4} \iint \mathrm{d}\nu_s \mathrm{d}\nu_i\, S_{sym}(\nu_s,\nu_i)  \nonumber \\
    &\times H(\omega_0+\nu_s)H(\omega_0+\nu_i) e^{-i(\nu_s+\nu_i)\tau}.
\end{align}

Let us note that the interferogram $M(\tau)$ incorporates quantum interference terms between pairs of pathways (postselected so that both photons emerge from BS$_1$ in mode $\hat{b}$), among i) through iv) in Fig.\ref{fig:schematic}(b). Thus,  $M_1(\tau)$, which does not depend on $\omega_0$, represents interference between the two unbunched pathways iii) and iv), and leads to essentially the same physics as in HOM interference.   This term yields information equivalent to a standard HOM-based QOCT apparatus.  $M_2(\tau)$, oscillating at frequency $\omega_0$, represents interference between each possible bunched pathway [i) and ii)] with each possible unbunched pathway [iii) and iv)], i.e. between the following pairs of pathways: i) and iii), i) and iv), ii) and iii), as well as ii) and iv).   Finally, $M_3(\tau)$, which oscillates at frequency $2 \omega_0$, represents interference between the bunched pathways i) and ii).

\begin{figure*}
	 \centering
	  \includegraphics[width=0.85\textwidth]{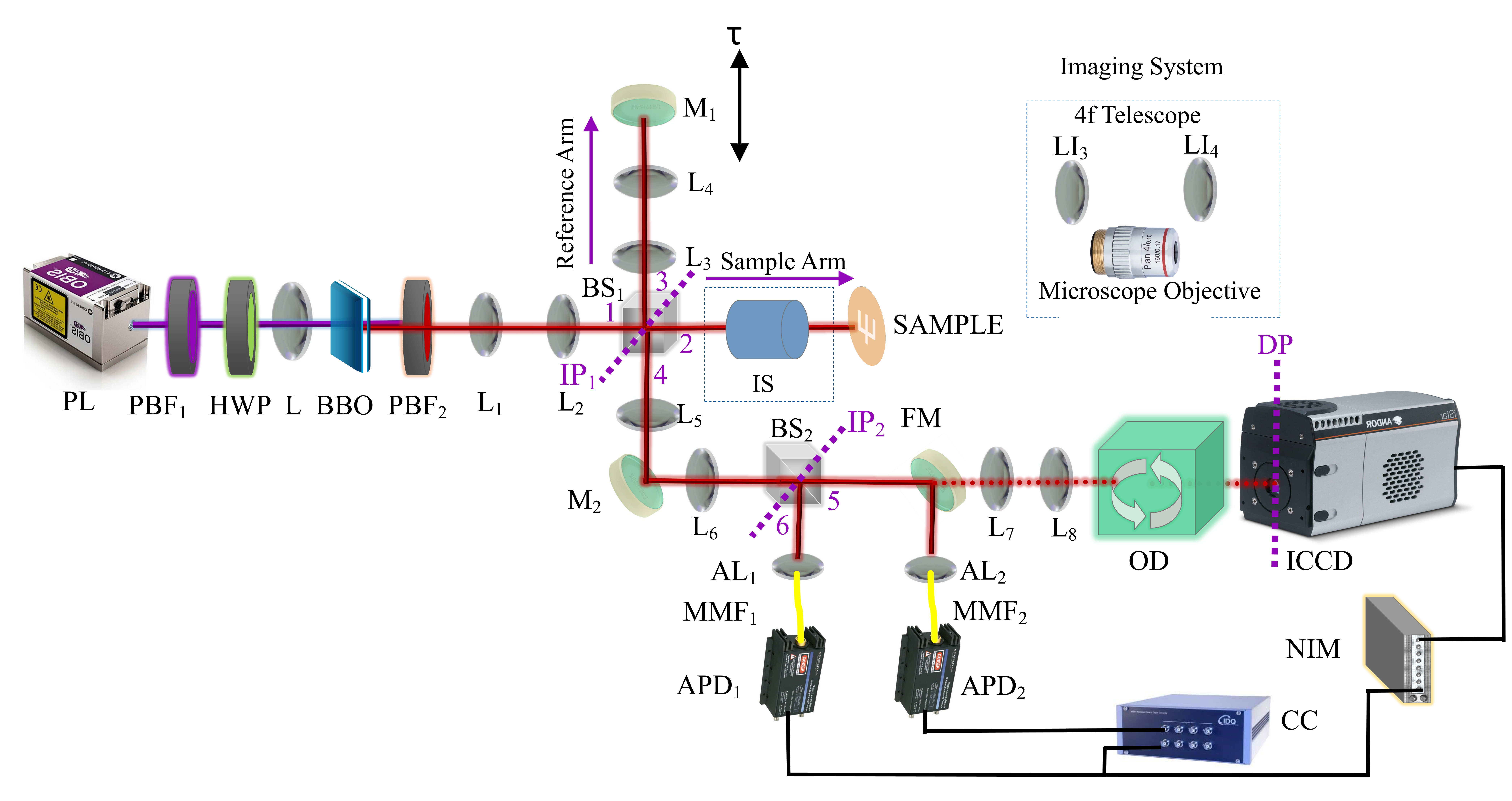}
	  	  \captionsetup{justification=justified}
	  \caption{Full-field QOCT / QOCM  experimental setup. PL:Pump Laser, PBF$_1$: Violet Bandpass filter, HWP: Half wave plate, L: Plano-convex spherical lens f=100 mm, BBO: $\beta$-Borate Barium Crystal, PBF$_2$: Infrared Bandpass filter and longpass filter, L$_{1-8}$: Near-Infrared plano-convex spherical lens, BS$_{1-2}$: Beam Splitter, IP$_{1-2}$: Interference plane, M$_{1-2}$: Near-Infrared mirror, Sample, IS: Imaging system, FM: Flip mirror, AL$_{1-2}$: Aspherical lens, MMF$_{1,2}$: Multimode fiber, APD$_{1-2}$: Avalanche photo-detector, OD: Optical Delay, CC: Coincidences counter, ICCD: Intensified Charge coupled device camera, DP: Detection plane}
	   \label{fig:setup}
\end{figure*}

To conclude this section, we remark on the importance of Eq. (\ref{Eq:Mtau}). The raw interferogram obtained by a Michelson-type QOCT experiment, a priori, does not contain the benefits of the original QOCT with a HOM configuration (i.e., resolution enhancement and dispersion compensation). Nevertheless, the HOM-type QOCT interferogram is in fact present in $M(\tau)$, particularly in the term $M_1(\tau)$. In fact, we can numerically extract $M_1(\tau)$ through Fourier analysis to recover the HOM interferogram,  with the advantage of using a collinear configuration with the aforementioned benefits of simplicity, robustness, as well as a larger photon-pair flux.


\section{Methods}
\label{sec:Methods}

\subsection{Experimental design}

Figure 2 shows a detailed schematic of our full-field QOCT apparatus, based on either a Michelson or Linnik interferometer.   Our SPDC source utilizes a $\beta$ barium borate (BBO) crystal of $2$ mm thickness, cut at $29.2 \deg$ for collinear, frequency-degenerate type-I phasematching. The crystal is pumped by the beam from a diode laser (PL), centered at $\lambda_p=405$ nm, with power of $100$ mW, and focused with a  plano-convex lens (L) of $1000$ mm focal length to a beam diameter of $w_0\approx 300$ $\mu$m at the crystal, to produce photon pairs centered at $\lambda_{s,i}=810$ nm. Note that the pump linewidth of $\sim1.4$ nm is sufficient to essentially fully suppress cross-interference effects (dips or peaks which may appear for each pair of interfaces); see Ref. \cite{Graciano2019} for a description of this effect and of how it may be mitigated.
The combination of a long-pass filter and a band-pass filter (PBF$_2$) suppresses the  pump beam following the BBO crystal. 
A $3\times$ $4f$ telescope (formed by lenses L$_1$ [$f=100$ mm] and L$_2$ [$f=300$ mm]) images the photons from the crystal plane to the beam splitter cube (BS1), producing a transverse mode with a diameter of about 1 mm. The transmitted, signal-mode photons travel toward the sample arm while the reflected idler-mode photons constitute the reference arm of the interferometer. A $1\times$ $4f$ telescope, assembled by lenses L3 and L4, both with focal length $f=75$ mm, is placed in the reference arm, prior to the retroreflector mirror M$_1$.

Our setup is designed so that the user may select between the Michelson and Linnik versions of the interferometer according to the sample to be studied, as controlled by the type of imaging system (IS) in the sample arm.   While for the Michelson, a $1\times$ $4f$ telescope (lenses LI$_3$ and LI$_4$, both with focal length $f=75$ mm) is used, for the Linnik this telescope is replaced by a $4\times$ microscope objective with a $0.1$ numerical aperture.    The effect of the microscope objective is two-fold: i) it leads to some magnification, improving the camera resolution, and ii) the collection solid angle implies that scattered k-vectors which deviate up to a certain degree from specular (i.e. normal) reflection can still be detected.   The IS maps the transverse area of interest of the photon-pair stream onto the sample (S) and back into the BS, which is ultimately acquired by the ICCD camera.

The photon pairs exit the interferometer through port $4$ of BS$_1$. A $1\times$ $4f$ telescope formed by lenses L$_5$ and L$_6$ (both with focal length $f=250$ mm) maps the interference plane (IP$_1$) to a second beam splitter (BS$_2$).  The modes emanating from the two output ports of BS$_2$ (5 and 6) are coupled by means of aspherical lenses AL$_1$ and AL$_2$ (both with focal length $f=8$ mm) to multi-mode fibers MMF$_1$ and MMF$_2$, leading to a pair of avalanche photodiodes APD$_1$ and APD$_2$. The electronic signals produced by the avalanche photodiodes are recorded by a coincidence counter (CC).

Our experimental setup is designed to permit straightforward switching between a standard (i.e. non-spatially-resolved), and full-field configurations.  For the standard configuration, 
the flip mirror (FM) remains in place. The resulting interferogram provides information about the axial location of each interface in the sample.  Such a measurement, which is reasonably fast ($\sim 5$ minutes per scan), allows us to determine the required travel range for the scanning mirror to be used for the full-field scans.  Note that the idler photon is coupled into a multimode fiber, which on the one hand boosts the detected photon-pair flux and on the other hand widens the transverse detection range of the signal photon at the ICCD camera.

For full-field scans, the flip mirror is disabled, to allow the signal photon to propagate through a $4\times$ $4f$ telescope (lenses L$_7$ and L$_8$ with focal lengths $120$ mm and $150$ mm) which relays the sample image onto the input plane of the image-preserving optical delay (OD) line. The OD is about $28$ m long ($\sim 90$ ns) and is designed to over-compensate the insertion delay time of the ICCD camera \cite{Aspden2013}, so that displacing the end-mirror in the the reference arm allows us to obtain the full portion of interest of the interferogram.

The ICCD camera is operated in a gated configuration, collecting photons on path $5$ in coincidence with their corresponding siblings on path $6$. Photons propagating through path $6$ are collected by multimode optical fiber MMF1, leading to APD1. The transistor-transistor logic (TTL) output pulses from APD1 are discriminated, modulated, and delayed by a series of nuclear instrumentation module (NIM) standard elements and then used to trigger the ICCD camera with a 10 ns coincidence time window.

In this work, we test our apparatus on two types of sample. First, we use synthetic, two-interface samples with a distinct transverse structure on each of the interfaces.   Second, we use biological samples (onion tissue and dragonfly wings)  with a $\sim 100$ nm copper/silver deposition to enhance the reflectivity.    While the first type of sample permits a three-dimensional reconstruction of both interfaces including the transverse structure of each interface, the second type of sample permits a three-dimensional topographical reconstruction of a single interface.  Note that while for synthetic samples, we use the Michelson interferometer, for the biological samples we switch to the Linnik interferometer.

\subsection{Synthetic samples preparation}
\label{subsec:synthetic}

The first type of sample used in our experiments consists of a $12$mm-diameter, $174$ $\mu$m-thickness borosilicate glass coverslip, with refractive index $n=1.51$ at $800$ nm, with a thin-film copper deposition applied on both sides.  The copper thickness to be applied to the two surfaces is calculated to result in $45\%$ and $80\%$ reflectivities.
Using laser micromachining, we have engraved the Greek letters $\varphi$ and $\Psi$ on the front and back coatings of the coverslip, respectively, by spatially-selective copper removal. The letter dimensions are  $\sim$500 $\mu$m$\times 400\mu$m.

\subsection{Biological samples preparation}
\label{subsec:biological}

Since for realistic applications, e.g. in clinical settings, a QOCT /QOCM apparatus would most probably be used to probe biological samples, in this paper we present data for such samples.   In particular, we use onion tissue and dragonfly wings coated with a thin layer of either copper or silver so as to enhance their reflectivity.   In the case of onion tissue, a copper film was deposited using an evaporator. In the case of the  dragonfly wing, a silver film is deposited using the sputtering technique.

\begin{figure*}[ht]
	 \centering
	  \includegraphics[width=1\textwidth]{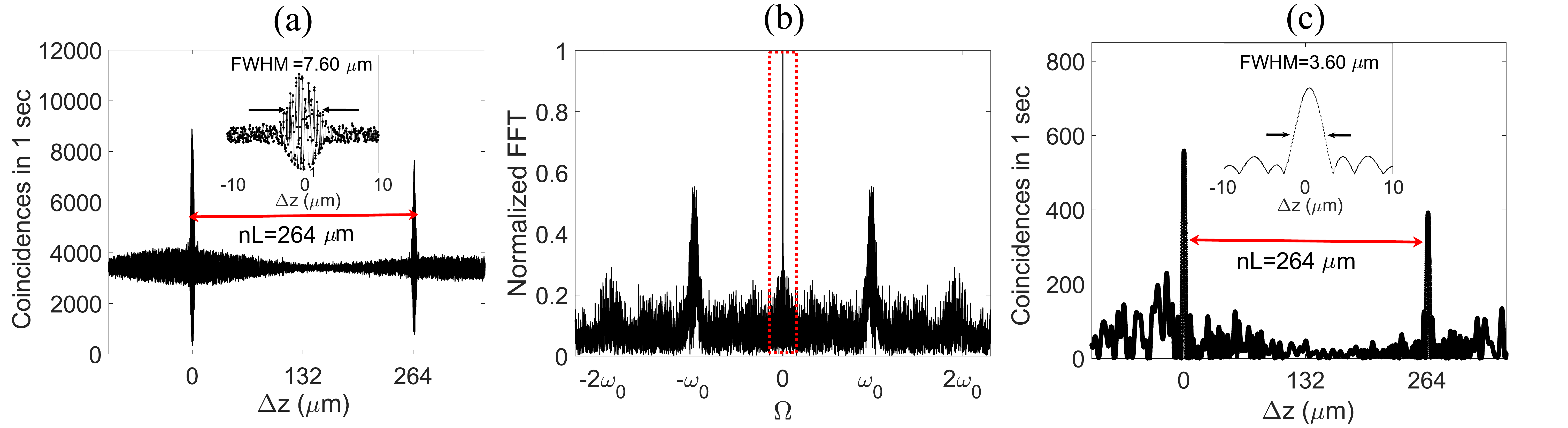}
	  	  \captionsetup{justification=justified}
	  \caption{Fourier filtering process. (a) Experimental two-photon interferogram  $M(\tau)$ obtained at the Michelson interferometer output, for unfiltered SPDC pairs (with 120 nm width).  Inset: closeup of left-hand peak.
	  (b) Fourier transform of $M(\tau)$, showing contributions at $\Omega=0$, $\Omega=\pm\omega_0$, and $\Omega=\pm 2\omega_0$.   
	  (c) Effect of  numerically filtering the $\Omega=0$ and carrying out an inverse Fourier transform to obtain $|M_1(\tau)|$, which is essentially equivalent to a HOM interferogram. Inset: close-up of left-hand peak.}
	   \label{fig:filter_algorithm}
\end{figure*}

\subsection{Fourier filtering algorithm}
Figure \ref{fig:filter_algorithm} describes the post-processing algorithm used to extract the QOCT interferogram, in particular applied to a non-spatially-resolved axial scan. In this case, the sample is a copper-coated coverslip of 174 $\mu$m thickness.  The photon pairs used are unfiltered, with  a $\sim120$ nm spectral width, while the interferogram is acquired with a $50$ nm  linear motor step.

The IS in the sample arm consists of a $1 \times$ $4f$ telescope. The FM remains in the raised position to reflect the photons and collect them in a multimode-mode optical fiber, to be detected by APD$_2$ (instead of by the ICCD camera). We acquire the spatially-unresolved coincidence detection rate as a function of the temporal delay $\tau$, defined by the position of mirror M$_1$ (see Fig. \ref{fig:setup}). This results in the raw coincidence interferogram shown in Fig. \ref{fig:filter_algorithm}(a), which corresponds to $M(\tau)$ (see Eq. \ref{eq:expansion}). In the inset, we show a closeup of the left-hand $M(\tau)$ peak.

We first subtract from the raw interferogram its mean, which is associated with the term $M_0$.   We then apply a fast Fourier transform (FFT) to the zero-mean interferogram, obtaining a trace as a function of $\Omega$, the Fourier conjugate variable to $\tau$; see Fig. \ref{fig:filter_algorithm}(b).  In this manner, we separate the three characteristic components ($M_1(\tau)$ centered at $\Omega=0$,  $M_2(\tau)$ centered at $\Omega=\pm\omega_0$, and $M_3(\tau)$ centered at $\Omega=\pm 2 \omega_0$).   Note that the $M_1(\tau)$ term leads to a single central peak, which contains information equivalent to a HOM interferogram and from which we can extract morphological information about the sample (see Ref. \cite{Lopez-Mago2012e}).   Subsequently, we apply a rectangular filter (see red dotted rectangle in Fig. \ref{fig:filter_algorithm}(b)) so as to only retain the central peak, followed by an inverse Fourier transform, from which we obtain an interferogram which is equivalent to that which would be obtained in a HOM measurement; see Fig. \ref{fig:filter_algorithm}(c).    Note that two main peaks appear each with a width of $\sim3.6$ $\mu$m, and each associated to one of the two interfaces; in the inset we show a closeup of the left-most $M_1(\tau)$ peak.  We see peaks instead of dips, because this measurement is based on coincidences at one of the exit ports of BS$_1$, instead of coincidences across the two outputs as in a standard HOM interferometer.  Note that an artifact associated with cross-interference between the two interfaces is not visible between the two main peaks, on account of the considerable pump bandwidth ($\sim 1.4$nm; see Ref. \cite{Graciano2019} for an explanation of this effect).    

Note also that the $M_3(\tau)$ peaks centered at $\Omega=\pm 2 \omega_0$ in Fig. \ref{fig:filter_algorithm}(b) have a limited amplitude.  Indeed, the oscillation at $2 \omega_0$  of these peaks  would require sampling at a finer step so as to fully resolve them.  This brings up the important point that since the sample's morphological information is extracted from the single $M_1(\tau)$ peak at  $\Omega=0$ (i.e. which does not oscillate), the linear step can be made larger, thus importantly reducing the acquisition time, even if $M_2(\tau)$ and $M_3(\tau)$ cannot be resolved.

\section{Experimental results}

\subsection{Full-field reconstruction: synthetic sample}

In this sub-section we present results obtained with our full-field QOCT setup, based on a collinear SPDC photon-pair source and on a Michelson interferometer.    This system is tested on a synthetic,  two-interface sample (with the Greek letters $\Psi$ and $\varphi$ engraved each on one of the surfaces; see section \ref{sec:Methods} subsection \ref{subsec:synthetic}).  In later sub-sections we present results obtained with our full-field QOCM setup which differs from the QOCT setup in the imaging system used in the sample arm, i.e.  a microscope objective is used instead of the a $1\times$ telescope.  We test this second system on metal-coated onion tissue and dragonfly wing samples.

\begin{figure*}[ht]
	 \centering
	  \includegraphics[width=0.8\textwidth]{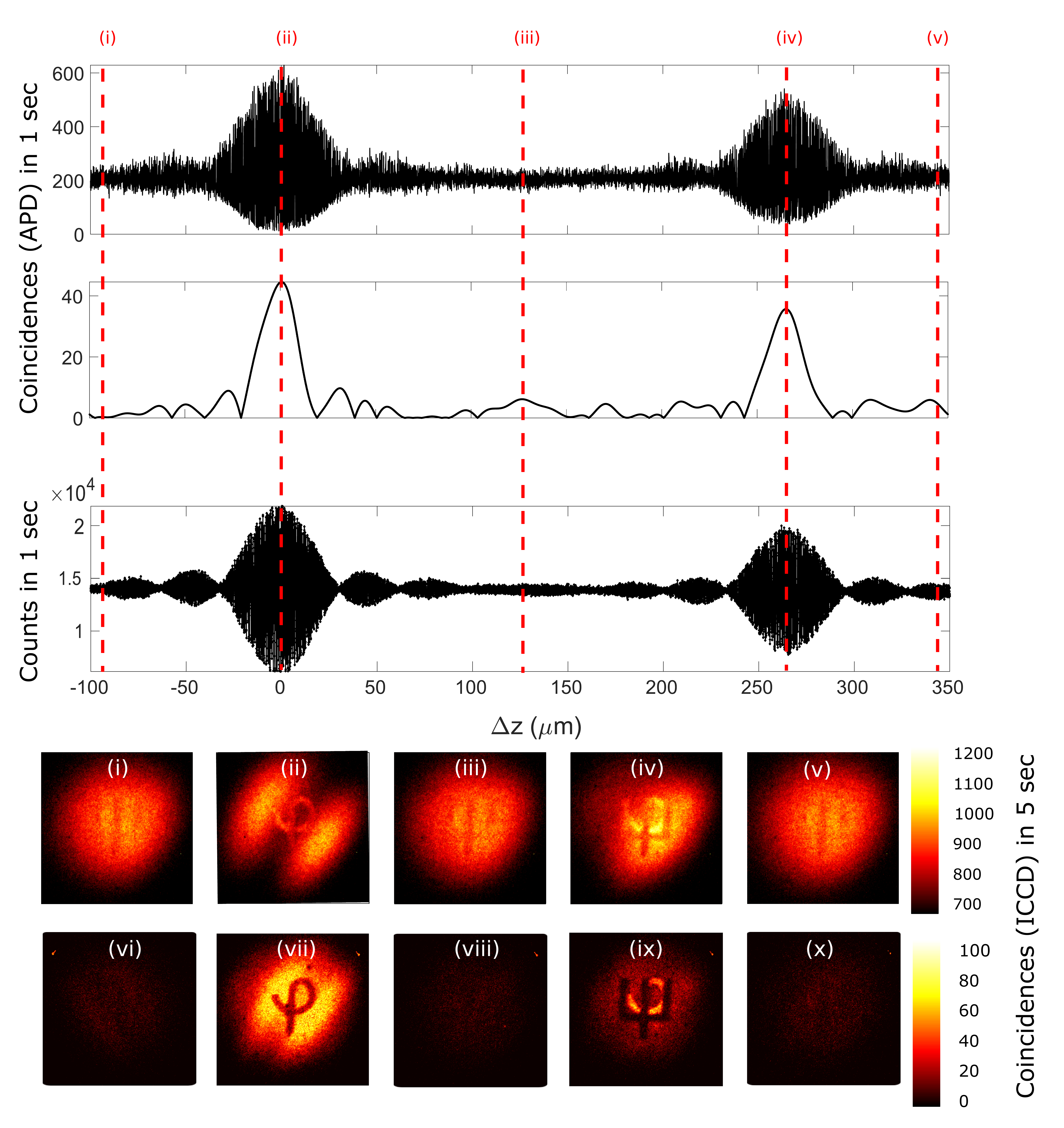}
	  	  \captionsetup{justification=justified}
	  \caption{First row: Raw interferogram $M(\tau)$ obtained for a synthetic sample [Cu-coated $174$ $\mu$m thickness coverslip with the letters $\varphi$ and $\psi$ engraved on the two surfaces]; $10$ nm-width spectral filter applied to the SPDC photon pairs.
	  Second row: Fourier extraction of the $\Omega=0$ contribution ($M_1(\tau)$).
	  Third row: OCT trace taken concurrently with the QOCT traces in the previous two rows.
	  Fourth row: Full-field interferogram obtained at the ICCD camera gated by each idler-photon detection at the five specific temporal delays identified by vertical red dotted lines in the first two rows.
	  Fifth row: Fourier-filtered interferograms at the same specific delay values, showing the  $\varphi$ and $\psi$ letters in high visibility only at those delays which corresponds to $|M_1(\tau)|$ peaks.}
	   \label{fig:full_field_recovery}
\end{figure*}

While the results in the preceding sub-section pertain to a single axial-only scan, here we move to a full-field configuration.  We use as sample a Cu-coated coverslip with engravings on both surfaces (see section 3 subsection \ref{subsec:synthetic} for a full description), preceded by a $1\times$ telescope as imaging system.   We lower the FM, see Fig. \ref{fig:setup},  so that the photon transmitted by BS$_2$ traverses a $4 \times$ telescope built from lenses L$_7$ and L$_8$ (with focal lengths $120$ mm and $150$ mm), an optical delay line, and reaches the ICCD camera; note that in this setup, plane IP$_2$ at BS$_2$ is imaged to the ICCD plane DP.  The photon reflected by BS$_2$ is detected by avalanche photodiode APD$_1$. We monitor the coincidence detection rate by triggering the ICCD camera using the electronic pulse generated for each detection event by APD$_1$ (see section \ref{sec:Methods}).

For an adequate understanding of the results to be presented here, it is useful to present a matching non-spatially resolved Michelson interferogram (we can straightforwardly enable or disable spatial resolution through the FM).  The first row of Fig. \ref{fig:full_field_recovery} shows the  interferogram for a non-spatially-resolved axial scan (similar to Fig. \ref{fig:filter_algorithm}(a), corresponding to $M(\tau)$), while the second row shows the Fourier filtered interferogram (i.e. corresponding to $M_1(\tau)$).   The third row shows the corresponding OCT trace, which is acquired concurrently with every QOCT measurement, and which is revealed simply by displaying single-channel rather than coincidence counts. Note that while the QOCT / QOCM measurement has important advantages, the OCT measurement yields a factor of $\sim 36$ greater counts as compared to the raw QOCT trace and a factor of $\sim 500$ greater than the Fourier-filtered QOCT trace.    The ability to yield concurrent OCT and QOCT measurements is an important advantage of the Michelson / Linnik approach in comparison with the HOM approach; indeed, in the HOM case the single-channel counts exhibit no delay dependence.  

There are five specific delay values indicated in the  first thee rows of Fig.\ref{fig:full_field_recovery}  as i), ii), iii), iv), and v).    Each of the five panels on the third row shows the coincidence detection rate, at each of the delay values i)-v). Note that in panels i), iii), and v) which corresponds to locations outside of the $M_1(\tau)$ interference peaks (see second row), the corresponding Greek letter is appreciable with very low visibility.    In contrast, panels ii) and iv) which correspond to the maxima of the $M_1(\tau)$ interference peaks, show a higher visibility.  

\begin{figure*}[ht]
	 \centering
	  \includegraphics[width=0.85\textwidth]{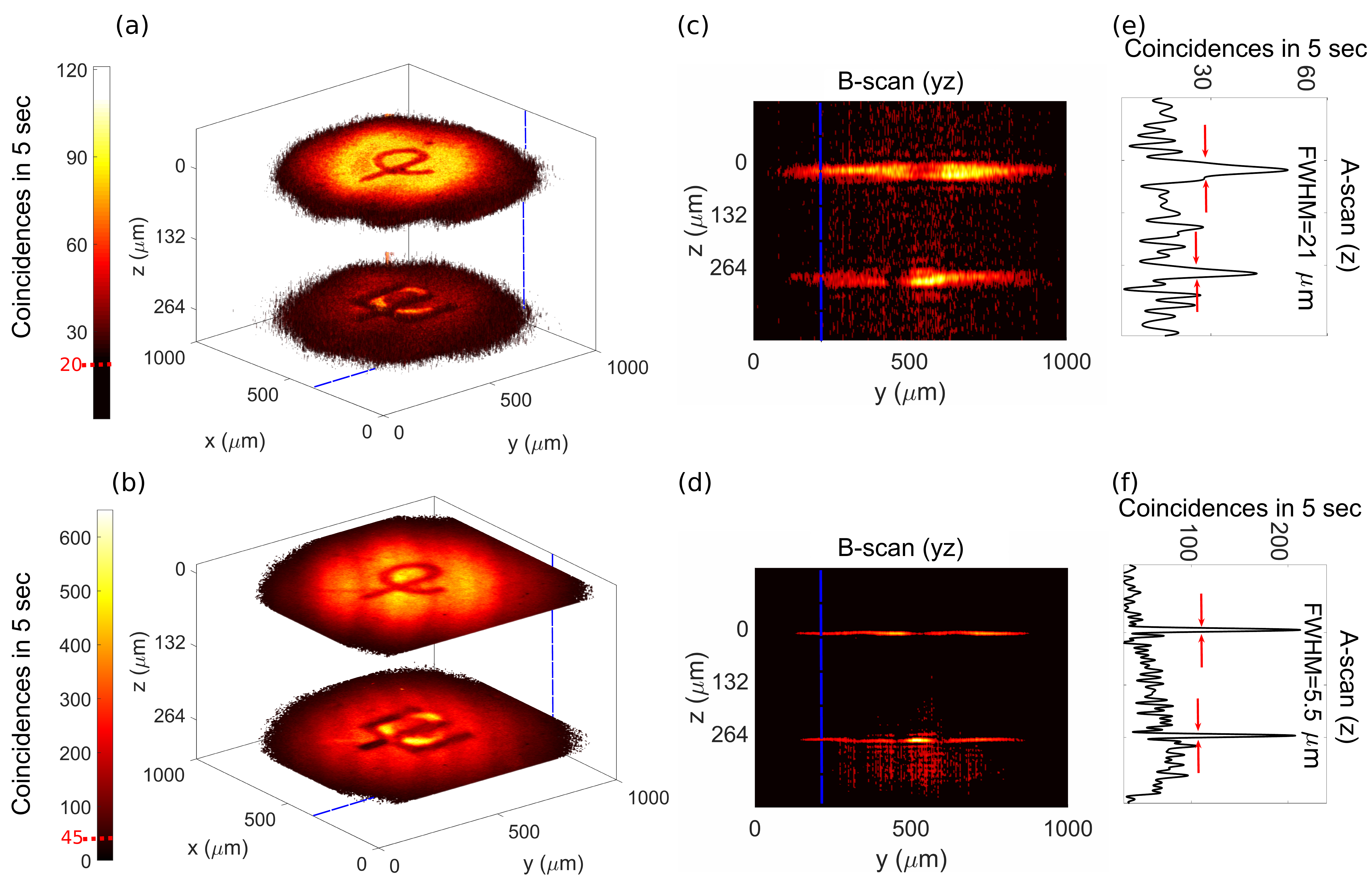}
	  	  \captionsetup{justification=justified}
	  \caption{Three-dimensional reconstruction of the synthetic sample. (a) Stack of all Fourier-filtered planes, for a $10$ nm SPDC filter width, plotted so that locations at which the signal does not exceed a certain threshold (indicated in red in the colorbar) are left unmarked. (b) Similar to panel (a), for the case of unfiltered photon pairs, with a $\sim120$ nm spectral width. (c) B-scan obtained from (a) for a fixed $x$ value [shown as blue line in (a)]. (d) B-scan obtained from (b) for a fixed $x$ value [shown as blue line in (b)].   (e) A-scan obtained from (c) for a fixed $y$ value [shown as blue line in (c)]. (f) A-scan obtained from (d) for a fixed $y$ value [shown as blue line in (d)].}
	   \label{fig:two_letter_tomography}
\end{figure*}
\nopagebreak

Note that while only five specific delay values are shown in the third row of Fig. \ref{fig:full_field_recovery}, we in fact acquired data for delay values spanning a total range of $600$ $\mu$m, with a $1$ $\mu$m ($0.5$ $\mu$m) step for photon pairs with a width of $10$ nm  ($120$ nm), i.e.  corresponding to $600$ ($1200$) delay values, and an integration time of $5$ s per full-field transverse acquisition sequence, resulting in a $50$ minute ($100$ minute) total acquisition time.  We have utilized a region of the ICCD camera of $300 \times 300$ pixels, so that our data is equivalent to $90,000$ individual axial scans, each with $600$ delay values.   The Fourier filtering scheme explained above (see section section \ref{sec:theo} and figure \ref{fig:filter_algorithm}) is then applied to each of these individual effective axial scans.   The resulting $M_1(\tau)$ traces can be put together and plotted as a function of the transverse coordinates for specific delay values of interest.  This is done for the five delay values i)-v)  indicated in the first two rows of Fig. \ref{fig:full_field_recovery}.   The results are shown in the fourth row of the same figure.   Interestingly, for delay values lying outside of the $M_1(\tau)$ peaks, the transverse intensity distribution show only some noise remnants, while delay values at $M_1(\tau)$ peak maxima show the Greek letters with an excellent visibility.

In order to visualize the full three-dimensional structure of the sample, Fig. \ref{fig:two_letter_tomography}(a) and (b) show stacks of all 600 Fourier-filtered planes, such as those in the third row of Fig. \ref{fig:full_field_recovery}, with all locations involving values not exceeding a certain threshold left unmarked.  Note that the specific  threshold values used are indicated in red in the colorbars.
This results in a hollow structure with the two interfaces showing their respective transverse spatially-resolved structures.   We have repeated this experiment for two experimental configurations: i)  spectral filter of $10$nm width applied to the photon pairs shown in the first row (panel a), and, ii) no spectral filter applied, with the full $\sim120$ nm photon pair bandwidth retained shown in the second row (panel b). We have also shown a B-scan (along plane XZ or YZ), for each of these configurations; see panel c) for configuration i), and panel d) for configuration ii).   Note that these B-scans correspond to a `slice' of the plots in panels a) and b) for a fixed $x$ value.  Additionally, we have presented in panels e) and f) A-scans  which correspond to the B-scans of panels c) and d), with a fixed value of the $y$ coordinate.    

Let us turn to the question of the resolution achieved, both axial and transverse.   While the integration time for the data shown in Fig. \ref{fig:filter_algorithm} (with an APD) was $1$ s per axial position, this time turned into $5$ s for the spatially-resolved data in Fig. \ref{fig:two_letter_tomography} (with the ICCD camera); on account of this, so as to maintain the total acquisition time manageable, the step size was adjusted from $50$ nm for the non-spatially resolved data to either $1$ $\mu$m or $0.5$ $\mu$m (depending on the SPDC spectral width) for the spatially-resolved data.  This implies that while the axial resolution (for the configuration without filter) apparent in Fig.\ref{fig:filter_algorithm}{\color{blue}(c)} is $3.6$ $\mu$m, in the B- and A-scans in Fig. \ref{fig:two_letter_tomography} it turns into $5.5$ $\mu$m (and $21$ $\mu$m for the case of a $10$nm bandpass filter). Note that the transverse resolution is limited by the pixel dimensions ($13\mu\mathrm{m} \times 13\mu$m) together with the total magnification of $4\times$ from the sample to the ICCD camera, resulting in a camera resolution of $\sim 7.5\mu$m (given by $2.3\ \times$ pixel size $/$ total magnification).

\begin{figure*}
	 \centering
	  \includegraphics[width=0.75\textwidth]{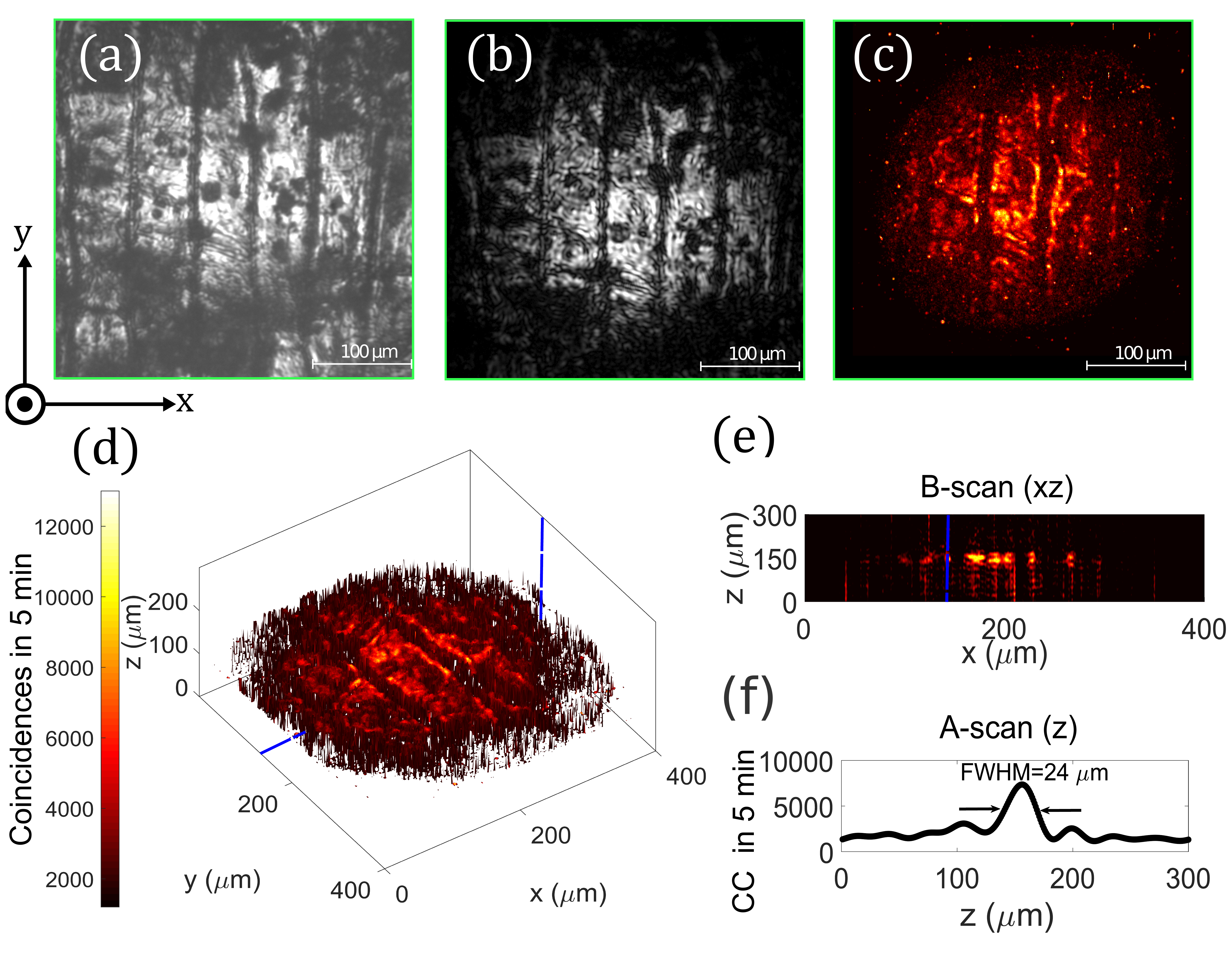}
	  	  \captionsetup{justification=justified}
	  \caption{Three-dimensional reconstruction of Cu-coated onion tissue.  (a) Transverse sample structure, as probed with white light. (b) Similar data obtained with a laser beam at $\lambda=810$ nm. (c) Transverse sample structure, as obtained with our QOCM apparatus, applying the Fourier filter, for a specific delay value. (d) Sample topography obtained from the stack of all Fourier-filtered transverse planes; while the $z$ value indicates the location at which the maximum counts are observed (for each combination of $x$ and $y$), the color indicates the value of the maximum.
	  (e) B-scan obtained from (d) for a fixed $y$ value [shown as blue line in (d)].
(f) A-scan obtained from (e) for a fixed $x$ value [shown as blue line in (e)].}
	  \label{fig:onion_tomography}
\end{figure*}

We note that our system is able to resolve well the structure (Greek letters $\psi$ and $\varphi$) on both interfaces.  We also note that while the letter $\varphi$  is clearly visible on the first interface, the letter $\psi$ is mainly visible on the second surface with the letter $\varphi$ overlapped.    This is due to the fact that while a photon probing the first surface interacts only with this first surface, a photon probing the second surface needs to make it through the first. In order to mitigate this effect and be able to observe each surface with its structure by itself, an optical system with a reduced depth of field could be used.  In our case, the telescope's depth of field for the signal photon is of the order of mm, clearly longer than the separation between the interfaces.    Note that in our QOCM measurements below, for which we use a microscope objective as imaging system, the Rayleigh distance reduced to $\sim 162$ $\mu$m.

\subsection{Full-field reconstruction: onion tissue and dragonfly wing}

An important step in the development of QOCT / QOCM technology is testing the performance of such systems on biological samples.  
Because the biological samples tend to have a low reflectivity in their natural state, we coat them with either copper or silver so as to make them highly reflective.    This implies that our QOCM apparatus can in principle recover the surface topography of the single coated interface, rather than the transverse morphology of more than one interface as was the case for the synthetic samples.  As mentioned before, in order to study the biological samples, we replace the $1\times$ telescope by a $4\times$ microscope objective with a numerical aperture of $0.1$.  The effect is two-fold: on the one hand this improves the transverse resolution, and on the other hand reflected photons with some deviation from specular (normal) reflection can still be collected and detected.  However, despite the use of a metal coating and of a microscope objective, reflection from the surface of the biological samples with a considerable degree of non-uniformity implies that a large fraction of the reflected photons will lie outside of the angular acceptance of the optical system, so that the available flux will be inevitably  sharply reduced.     This in turn translates into the need for longer acquisition times.

The biological sample in question is affixed to a glass substrate and placed as end mirror in the sample arm of the interferometer.  We probe the sample with three
distinct illumination systems: i) a white light source, ii) a diode laser at $810$nm, as well as iii) through our QOCM apparatus.  In the case of i) and ii), a flip mirror prior to the crystal is raised so as to send the photons from the illumination system along the same path as the collinear SPDC photon pairs in iii), while we block the reference arm and operate the ICCD camera in ungated mode.   Note that when using the white light source or the diode laser, we use an iris to restrict the transverse dimensions of the beam to approximately the same as for the SPDC photon pairs.  The measurements with illumination systems i) and ii) serve as a useful reference against which to compare our QOCM measurements.  

\begin{figure*}
	 \centering
	  \includegraphics[width=0.75\textwidth]{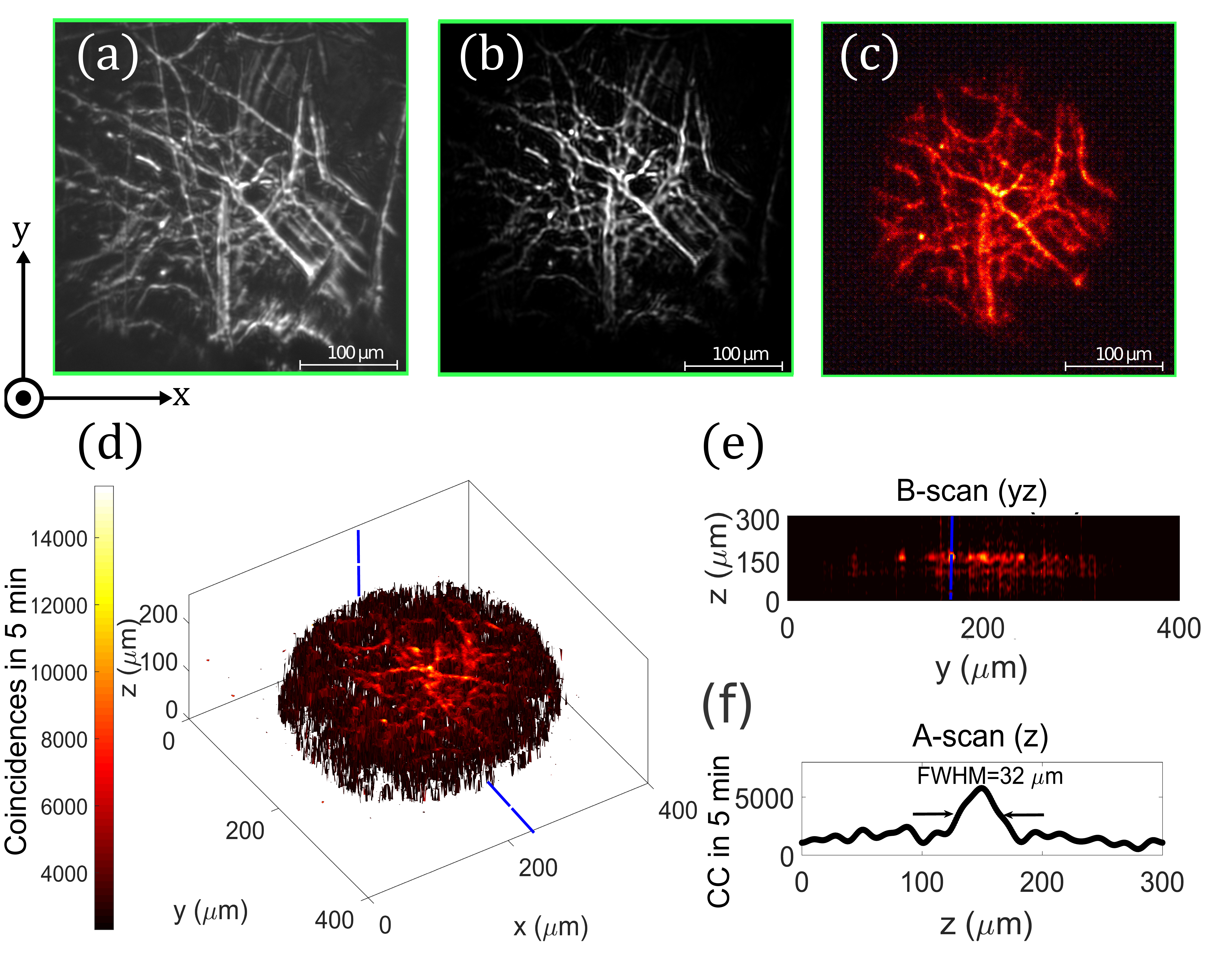}
	  \captionsetup{justification=justified}
	  \caption{Three-dimensional reconstruction of Ag-coated dragonfly wing.  (a) Transverse sample structure, as probed with  white light.  (b)   Similar data obtained with a laser beam at $\lambda=810$ nm.  (c) Transverse sample structure, as obtained with our QOCM apparatus,  applying the Fourier filter, for a specific delay value.   (d) Sample topography obtained from the stack of all Fourier-filtered transverse planes; while the $z$ value indicates the location at which the maximum counts are observed (for each combination of $x$ and $y$), the color indicates the value of the maximum.
	  (e) B-scan obtained from (d) for a fixed $x$ value [shown as blue line in (d)].
(f) A-scan obtained from (e) for a fixed $y$ value [shown as blue line in (e)]}
	  \label{fig:dragonfly_wing_tomography}
\end{figure*}

In our QOCM system, the transverse region in the sample which can be probed has an area of $\sim 360 \mu$m$\times 360\mu$m, corresponding to $400 \times 400$ pixels on the ICCD sensor.  We scan the delay over a range of $300$ $\mu$m, with steps of $1\mu$m, with an integration time per delay value of $5$ minutes (while $5$ s were used for the case of the synthetic samples).   While the total acquisition time of 25 hours is certainly large, we note that this corresponds to $160,000$ effective axial scans each with $300$ delay stops. 
Importantly, note that the data acquisition speed-up $\xi$ made possible by full-field / multi-mode detection of the signal /idler photon  is given by  $\xi=N^2 \xi_{ax}$ where $N^2$ is the number of pixels used (in this case $N^2=160,000$), and where $\xi_{ax}$ represents the speed-up per axial scan resulting from the collinear / multi-mode design of our setup.
In other words, were the reconstruction of the sample to be carried out through a transversely-rasterized sequence of axial scans, the time required for the measurement would be greater by a factor $\xi$.
Note that rasterizing at one transverse point per minimum resolvable distance would still give a $>5\times 10^3$  data acquisition speed-up in our case (this speed-up would be evidently be greater for systems with a higher transverse resolution).

Note that while the transverse resolution (limited by the diffraction limit of the microscope objective) is 
$\sim 4.9$ $\mu$m, the axial resolution is dependent on the SPDC bandwidth and is similar to that obtained for the synthetic samples.

Fig. \ref{fig:onion_tomography} shows our experimental data for the Cu-coated onion tissue, for an SPDC spectral width of $10$nm.   While panel a) shows our data obtained for the white light source, panel b) shows our data obtained for the diode laser.    Panel c) shows a C-scan (i.e corresponding to a xy plane with a fixed z value) obtained with the full QOCM setup at a certain fixed delay.  Note that these three panels, amongst which a) and b) act as references, and c) shows an example of a QOCM measurement, agree well with one another.   Panel d) shows information from all 300 planes obtained for the 300 delay stops in the full measurement, with the $z$ coordinate indicating the specific delay stop at which the count rate is maximized for a given choice of $x$ and $y$ coordinates, and where the color indicates the value of this maximum.   In this manner, panel d) depicts the topography of the single coated interface.  Cell structures and cell junctions can be clearly seen.   Panel e) shows information from all 300 planes in the form of a B-scan, i.e corresponding to a $xz$ plane with a fixed $y$ value (indicated by blue lines in panel d).   Panel (f) shows an A-scan obtained from the B-scan of panel e) by fixing the value of x (indicated by a blue line in panel e).   Note that the width of the A-scan peak is consistent with that observed (for a spectral filter of $10$ nm width applied to the photon pairs) in the case of the synthetic sample.   

Figure \ref{fig:dragonfly_wing_tomography} is similar to Fig. \ref{fig:onion_tomography}, for an Ag-coated dragonfly wing.   A dragonfly wing consists of two main structures: the veins and the membrane, according to Sun \textit{et al.} \cite{Sun2012}. The membrane consists of a waxy layer that shows different arrangements of lines. According to  Kreusz \textit{et al.} \cite{Kreuz2001},  this membranous zone consists of a cuticle composed of two chemical components: chitin embedded in the form of fibrillations in a matrix containing numerous structural proteins and lipids and a thin waxy layer.  In our measurements, we have probed a membranous area, which does not encompass any of the main veins. 

In a different, recently-published approach to quantum microscopy
one of the photons is transmitted through a transparent, single-layer sample (i.e. limited to two interfaces), with a transversely-varying width.   Making use of the one-to-one correspondence between the coincidence count level on one of the edges of the Hong-Ou-Mandel dip and the optical thickness, it becomes possible to recover the optical thickness, with  sub-micron resolution, as a function of the transverse position \cite{Ndagano2022}.  We note that while in our own work we regard the entire width of the relevant interferogram peak as the axial resolution, we could take a similar approach as in the above paper to map the transversely-varying coincidence count level, at a fixed delay corresponding to a given interface, to transversely-dependent variations in the axial position of the interface.  While this would boost the axial resolution (with narrower interferogram peaks yielding  higher resolutions), such an approach would hinge on assuming that any coincidence rate variation is due to quantum interference, i.e.
disregarding transverse variations in reflectivity, absorption, and scattering, which likely occur in biological samples.

\section{Conclusions}

In this work we have introduced and demonstrated a quantum optical coherence microscopy (QOCM) technique.  Our work is based on previous QOCT implementations, and incorporates a number of key features: i) it relies on a collinear SPDC photon-pair source, which means that in principle all emitted photon pairs may be used, ii) consistently with such a collinear design, quantum interference takes place in a Michelson / Linnik interferometer instead of the more typical Hong-Ou-Mandel, iii) it incorporates multi-mode collection for the idler photon and full-field detection for the signal photon, thus boosting the usable flux without harming the quantum interference and likewise maximizing the transverse spatial extent of the sample which can be resolved, iv) an SPDC pump bandwidth is used which is large enough so as to nearly fully suppress artifacts due to cross-interference, and v) our experimental design permits the concurrent acquisition of  OCT and QOCT traces, the former with a grater flux and the latter with quantum-conferred advantages. 

We have tested our apparatus on representative samples including a synthetic Cu-coated fused silica substrate, with a design engraved on both surfaces, as well as metal-coated biological samples, namely onion tissue and a dragonfly wing.   Our system achieves $\mu\mathrm{m}$-level axial and transverse resolutions, with the benefit of quantum-conferred advantages such as factor of 2 enhancement of the axial resolution and even-order dispersion cancellation.  Importantly, our full-field detection implies a speed-up in data acquisition with respect to an equivalent transversely-rasterized axial scan approach equal to the number of pixels used (in our case equal to $1.6\times 10^5$). While the collection times (hours) are still not short enough for clinical use, we believe that this work represents a significant step forward.

\begin{acknowledgments}
This work was funded by the Consejo Nacional de Ciencia y Tecnolog\'{i}a (CF-2019-217559); PAPIIT-UNAM (IN103521) and AFOSR (FA9550-21-1-0147).
\end{acknowledgments}

\bibliography{bibliography}

\end{document}